# Assessing Graph-based Deep Learning Models for Predicting Flash Point

Xiaoyu Sun,* Nathaniel J. Krakauer,* Alexander Politowicz, Wei-Ting Chen, Qiying Li, Zuoyi Li, Xianjia Shao, Alfred Sunaryo, Mingren Shen, James Wang, Dane Morgan[+]

[*]     Equal contributions

[+]     Corresponding author – ddmorgan@wisc.edu, Dept. of Materials Science and Engineering, 244 MSE, University of Wisconsin, Madison 53562

Flash points of organic molecules play an important role in preventing flammability hazards and large databases of measured values exist, although millions of compounds remain unmeasured. To rapidly extend existing data to new compounds many researchers have used quantitative structure-property relationship (QSPR) analysis to effectively predict flash points. In recent years graph-based deep learning (GBDL) has emerged as a powerful alternative method to traditional QSPR. In this paper, GBDL models were implemented in predicting flash point for the first time. We assessed the performance of two GBDL models, message-passing neural network (MPNN) and graph convolutional neural network (GCNN), by comparing against 12 previous QSPR studies using more traditional methods. Our result shows that MPNN both outperforms GCNN and yields slightly worse but comparable performance with previous QSPR studies. The average $R^2$ and Mean Absolute Error (MAE) scores of MPNN are, respectively, 2.3% lower and 2.0K higher than previous comparable studies. To further explore GBDL models, we collected the largest flash point dataset to date, which contains 10575 unique molecules. The optimized MPNN gives a test data $R^2$ of 0.803 and MAE of 17.8 K on the complete dataset. We also extracted 5 datasets from our integrated dataset based on molecular types (acids, organometallics, organogermaniums, organosilicons, and organotins) and explore the quality of the model in these classes.

## 1 Introduction

Quantitative structure–property relationship (QSPR) analysis has been widely used in predicting various properties of molecular systems. As a popular analytical method, many previous studies have been done on assessing QSPR.[1--4] QSPR uses features, typically directly calculated from molecular structure (e.g., bond networks, functional groups, molecular size) but also potentially from measured properties (e.g., boiling points), and correlates them with measured properties of interest, such as toxicity, solubility, and flash point. The correlations are then applied to predict the target properties for new systems. A large list of features has been developed through extensive human ingenuity and testing over many decades, and now thousands of possible features can be determined from just the basic molecular description (e.g., SMILES string).

In recent years, with the development of deep learning, machine learning models have become increasingly popular as a QSPR method. In particular, graph-based deep learning (GBDL) models have emerged as a promising QSPR method in predicting properties of molecular systems.[6-11] GBDL models map molecules to graphs, where nodes are atoms and edges are chemical bonds. These graphs are represented by adjacency matrices. Features, such as atom features, bond features and geometrical features, are also represented as matrices, which are then incorporated together with adjacency matrices by using various algorithms and model architectures to predict molecular properties.

Traditional QSPR features are motivated by chemical intuition and therefore have had significant human input into their development. GBDL methods use a limited set of fundamental features specifying the basic atomic structure of the molecule. The GBDL approaches may therefore offer more flexibility in systems with novel chemistry and potentially provide greater accuracy on problems where the deep learning algorithms can discover better feature maps than those developed by humans. It is therefore important to assess how well such GBDL methods perform compared to more traditional QSPR approaches based on more human engineered features.

In this work we focus on modeling flash points. This physical property was chosen for the following three reasons: First, flash point is an important property to prevent fire hazard related to the storage, transport, and use of flammable substances, so accurate models are of value. Second, previous studies and public chemistry databases provide enough flash point values to train deep learning models and provide robust model assessment. To enable this study, we have collected the largest flash point dataset to date of which we are aware, which contains 10575 unique molecules, all with valid SMILES strings (see section 2.1 for details of the database). Third, to the best of our knowledge, no researchers have used GBDL to predict flash point in the literature, so a comparison between GBDL models and traditional QSPR approaches in predicting flash point provides a test of the graph-based methods for a new domain of significant practical importance.

This work assesses the performance of GBDL models in predicting flash points with two methods. First, by comparing our results with previous studies that use traditional QSPR approaches, and second, by using our models to predict a test sample of our dataset as well as samples of data in different chemical domains. We apply two GBDL models that are implemented in DeepChem:[13] Graph Convolutional Neural Network (GCNN)[8] and Message Passing Neural Network (MPNN).[11] To conduct an effective comparison to previous QSPR studies, we first identified the datasets and data partitioning methods from each previous study, then trained and tested optimized GBDL models using these datasets.

The samples from the comprehensive dataset and chemical domain datasets were used to measure the robustness and chemical domain adaptability of the GBDL models. Fitting and testing on our complete distribution assessed the ability of our models to predict flash points across many chemistries. Training on the full dataset and assessing the fit accuracy on specific chemical subsets assessed how well our models predict flash point in specific domains of potential interest, which may be significantly different from the average accuracy on the total dataset. For subsets we focused on acids and different types of organometallics. The motivation for these subsets is that they represent categories with different chemistry, e.g., governed by their metal content, which might make them both of interest for certain applications and susceptible to significant fitting bias. For example, acids are of interest for their corrosive properties and represent a certain health hazard, and organisilicons are of interest as sealants and herbicide additives. The studied categories are representative examples of how a flash point model might be used in a specific domain, although these are by no means the only possible examples.

The paper in organized as follows. Sec. 2 describes the methods, including properties of all the datasets and models. Sec. 3 gives the results, including a comparison of the GBDL models to previous studies (Sec. 3.1), an assessment of the GBDL models on the total dataset (Sec. 3.2), and an assessment of the GBDL models on the specific chemically related subsets (Sec. 3.3). Sec. 4 provides a summary of conclusions and Sec. 5 provides a summary of the data and code management.

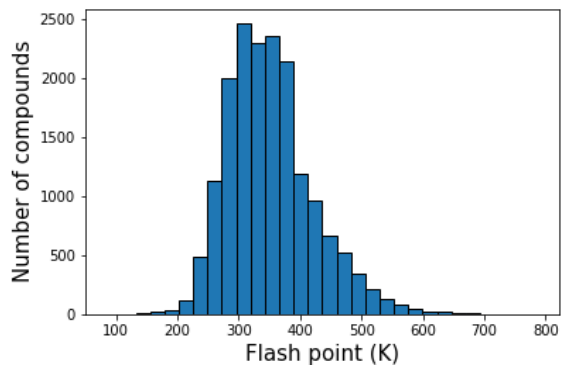

**Figure 1**. Flash point distribution of integrated dataset after removing duplicates.

| | Chen 14 | Wang 11 | Mathieu 12 | Mathieu 14 | Katrizky 07 | Saldana 11 | Le 15 | Pan 07 | Pan 13 | Patel 10 | Carroll 10 | Carroll 11 | Carroll 15 | Godinho 11 |
|---|---|---|---|---|---|---|---|---|---|---|---|---|---|---|
| Mean (K) | 315.5 | 314.9 | 314.9 | 338.6 | 308.2 | 324.5 | 371.4 | 287.1 | 313.6 | 328.4 | 293.1 | 312.7 | 324.1 | 334.6 |
| Stdev | 46.1 | 44.6 | 44.6 | 69.6 | 48.6 | 56.2 | 74.1 | 37.5 | 44.1 | 64.9 | 39.1 | 41.4 | 39.4 | 18.9 |
| Size | 211 | 210 | 210 | 1328 | 739 | 631 | 8843 | 92 | 193 | 234 | 292 | 937 | 81 | 75 |
| Original Size | 236 | 230 | 230 | 1457 | 758 | 631 | 9399 | 92 | 207 | 236 | 300 | 1000 | 82 | 103 |

**Table 1.** Statistics of extracted datasets from papers. Names refer to first author and year for the following references: Chen14 [19], Wang11 [14], Mathieu12 [23], Mathieu14[27], Katrizky07 [28], Saldana11 [26], Le15 [21], Pan07 [22], Pan13 [24], Patel10 [25], Carroll10 [16], Carroll11 [15], Carroll15 [18], Goinho11 [20]. "Original size" is the size of the dataset taken from the reference. "Size" is the size of the dataset we used after removing cases where we could not obtain valid SMILES strings.

## 2 Methods

### 2.1 Dataset Description

We collected data from academic papers,[14--28] the Gelest chemical catalogue,[30] the DIPPR database,[12] Lange's Handbook of Chemistry,[31] the Hazardous Chemicals Handbook,[32] and the PubChem chemical database.[29] Basic properties of the flash point datasets from each paper (size, average, and standard deviation) are given in Table1. Only experimental data was used. In particular, only flash point values that were stated to be experimental were collected from DIPPR, which also contains significant amounts of flash point data predicted by models. In the collection process, we included flash point values measured with open cup and closed cup methods together, although some subsets from specific papers used only closed cup. The advantage of including both types of measurements is that this provides the largest possible dataset for training. As stated by Le et al.,[21] the closed cup flash point is typically 5-10 K lower than the open cup one and experimental errors are typically 5-8 K. Thus the errors between methods are close to the inherent experimental errors and we feel that little accuracy is lost in the machine learning modeling by including both open and closed cup measurements together. All distinct databases and datasets of which we were aware that were easily accessible and from the past ten years were collected. We excluded data in PDF formats that Tabula,[33] an open-source table extractor for PDF files, could not extract. We also excluded entries for which we could not obtain a valid SMILES string (see section 2.2). All collected data was integrated into a dataset of 17333 total and 10575 unique compounds. We remove all duplicates from the dataset as described in section 2.2, where duplicate is defined as having the same canonical SMILES representation. When duplicates occur, we find their mean and standard deviation. If the standard deviation of duplicate flash point values is greater than 5 K we remove all entries of that compound and take it to be uncertain. If the standard deviation is less than or equal to 5 K we keep one entry representing all the duplicates and assign it the mean value of all duplicate flash points. The distribution of flash points after removing duplicates is shown in Figure 1, and statistics on the whole dataset are given in Table 2. All the data is provided in the supporting information except data from DIPPR, as this database is proprietary.

We explored the accuracy of our models on specific chemical subsets of the data to determine if there are any biases related to specific chemistries. For the chemistry types we divided our entire dataset into five categories: acids (contain the word "acid(s)" in the chemical name), all organometallics (contain a metal or metalloid atom), organogermaniums (contain Ge), organosilicons (contain Si), and organotins (contain Sn). The size of each chemical dataset and their average and standard deviation flash points values are given in Table 2.

|  | Full Dataset | Acids | Organo-metallics | Organo-germaniums | Organo-silicons | Organo-tins |
|---|---|---|---|---|---|---|
| Mean (K) | 371.0 | 401.8 | 320.1 | 323.2 | 314.9 | 387.9 |
| Std | 73.3 | 65.8 | 52.3 | 37.9 | 43.5 | 78.9 |
| Size | 10575 | 1386 | 1576 | 35 | 1382 | 50 |

**Table 2.** Statistics for whole dataset and each chemical dataset.

**2.2 Data Preparation**

We collected compounds' IUPAC name and used an open source parser, OPSIN,[34] to generate SMILES strings from the IUPAC name. RDKit,[35] an open-source cheminformatics software package, was then used to canonicalize SMILES strings and validate their correctness. We removed compound names that OPSIN could not convert to SMILES strings, but the tool effectively generated SMILES strings for most of the chemical compounds. For any given test described below we removed duplicate canonical SMILES strings as appropriate so that training, validation, and test datasets never contained duplicates within or between each other. The final data refinement step was to keep only entries that were organic (and metalorganic) molecules. Specifically, we manually checked data extracted from non-academic paper sources, removing all non-organic molecules. We defined organic molecules simply as molecular systems having at least one carbon atom. Open-source software from the DeepChem library converted the SMILES strings to feature vectors used to train graph-based deep learning models. DeepChem's convolutional featurizer generator made features for the GCNN model and the weave featurizer generated features for the MPNN model.

**2.3 Models**

Implementations of MPNN and GCNN models were taken from the DeepChem library.[13] At a high level, both approaches abstract a molecule as a graph where the nodes are atoms and the edges are bonds. MPNN and GCNN features include atom degree, implicit valence, formal charge, number of radical electrons, and hybridization encoded via one hot encoding. Both models use a canonical adjacency list and bond list for graph representation. The MPNN model has an additional set of pair features, which encodes the connectivity of atoms per molecule.

Graph convolution models take the aforementioned features as initial features for every molecule, which features are then passed to graph convolution modules. In graph convolution modules, each of which contains a graph convolutional layer followed by a batch normalization layer and a graph pool layer, atom features are updated by combining with neighboring atoms using an adaptive function. Then feature vectors are passed to a fully-connected layer to finish the convolution. Finally, the feature vectors for all atoms are summed, generating a graph feature vector, which is fed to a regression layer for making predictions.[8] Our GCNN implementation[36] followed DeepChem's GCNN implementation, which contains two graph convolution modules followed by a fully-connected dense layer and a regression layer.

MPNN operates on undirected graphs with a feature vector per node and edge. The two forward feeding phases are a readout phase and a message passing phase. The readout phase computes the feature vector for the entire graph and the message passing phase updates hidden states at each node based on the selected update function. Our implementation of the MPNN model followed DeepChem's MPNN implementation,[36] which uses the best approaches for MPNN models determined in Gilmer et al.[11] A seq2seq[5] model implements the readout phase, which is a machine learning model used to abstract features from sets. An edge dependent neural network implements the message passing phase, which maps all neighboring atoms' feature vectors to updated messages merged by gated recurrent units. The seq2seq model implements the gated recurrent units. In the final readout phase all atoms' feature vectors are abstracted as a set. The seq2seq model updates the final feature vectors and outputs them to the graph representing the molecule.[19]

**2.4 Optimization and Evaluation of Graph-based Deep Learning Models**

We made our comparisons according to the metrics of Root Mean Squared Error (RMSE), Mean Absolute Error (MAE), R-squared Score ($R^2$), and Average Absolute Relative Error (AARE), although not all statistics were available from each paper. We used built-in methods in the DeepChem package[13] to calculate RMSE, MAE, and $R^2$. Equations used to calculate RMSE, MAE, $R^2$ and AARE are shown in Equation (1)--(4) respectively.

$$\sqrt{\sum_{i=1}^{N} \frac{(FP_{pred,i} - FP_{exp,i})^2}{N}} \quad (1)$$

$$\sum_{i=1}^{N} \left( \frac{|FP_{pred,i} - FP_{exp,i}|}{N} \right) \quad (2)$$

$$1 - \sum_{i=1}^{N} \frac{(FP_{pred,i} - FP_{exp,i})^2}{(FP_{exp,i} - \overline{FP_{exp}})^2} \quad (3)$$

$$\frac{100}{N} \sum_{i=1}^{N} \left( \frac{FP_{pred,i} - FP_{exp,i}}{FP_{exp,i}} \right) \% \quad (4)$$

* N is the number of compounds; $FP_{pred,i}$ and $FP_{exp,i}$ are, respectively, the $i^{th}$ molecule's predicted flash point value and experimental flash point value. $\overline{FP_{exp}}$ is the mean value of experimental flash point values

When doing literature comparisons, a nested 5-fold cross validation method was implemented for model optimization and error assessment. Comparisons were made to the best model from each paper. The primary papers that we compared against can be divided into two categories with respect to their testing: The first category used an 80 % training 20 % test splitting method ("80/20 split" papers) without providing the exact test set[15-17, 19-22, 24, 28]. Note that we used Katrziky07's ANN dataset partitioning method[28] as we compared to their ANN modeling. The second category provided a specific test set in the paper ("explicit test" papers),[14,18,23,25] and for this category we use the provided test sets for assessing our models. For the 80/20 split papers we used a 5-fold cross validation on randomly shuffled data, which should be a comparable model assessment to that used in the paper. The first fold was taken as the test set and the remaining 4 folds were combined as a training set. An optimization of the GCNN and MPNN model hyperparameters was then done on the training set by minimizing the (nested) 5-fold cross validation RMSE (details of the hyperparameter optimization are below). Then we trained our optimized model on the full training set, tested on the test set, and recorded test set RMSE, MAE, $R^2$ and AARE scores. To ensure stable error estimates for the test set, we repeated the aforementioned procedure until all 5 folds were used. The one exception to this approach was our analysis of Le15, where, due to the large size of the data, only one fold was used. It should be noted that one test fold of the Le15 dataset contains 1768 compounds, which size we believe ensures the robustness of test set prediction statistics. If the optimization-testing procedure was conducted more than once, we calculated the mean value and standard deviation for each metric as our final result. For the "explicit test" papers we optimized each GCNN model and MPNN model on the provided training set by minimizing the 5-fold cross validation RMSE. Then we trained our optimized model on the provided training set, tested on the provided test set, and recorded test set RMSE, MAE, $R^2$ and AARE scores as our final results. With these steps, we believe our comparisons are as accurate as possible. We note that some of the papers using boiling-point as a feature used other testing methods but as these papers are not appropriate comparisons for the GBDL methods (see Sec. 3.1) we do not make any attempt to mimic their testing approaches.

We tested the performance of the models on test sets constructed from the entire dataset, as well as from subsets broken out by specific chemical types, as described above. For the chemical subset test sets, we randomly selected 20% of acids, 20% of organometallics, 50% of organogermaniums, 20% of organosilicons, and 50% of organotins as test sets for each chemical type, respectively. It should be noted that, while we generally chose 20% size test datasets in this work when a choice was needed, the 50% of organogermaniums and organotins are used here to try to have a larger test dataset size as these chemical classes have relatively few members. For testing on the entire dataset, we integrated the aforementioned chemical test sets and randomly added points from the full database (excluding the chemical test sets) to the integrated test set to make the total test set size to be around 20% of the entire dataset size. Then all other points in our entire dataset were used as the training set. Before testing, we optimized the GCNN model and MPNN model hyperparameters on the training set by randomly choosing 20% of the training set as a validation set. For these tests, we didn't use multiple leave-out validation sets because the validation set is large enough to obtain a stable result and multiple hyperparameter optimizations on such a large training set was too computationally demanding. After optimizing, we trained our model on the training set, tested our model on total and chemistry specific test sets, and recorded test set RMSE, MAE, $R^2$ and AARE scores. Both the GCNN and the MPNN were optimized and trained on CPU and GPU architectures.

We used a simple grid search method to do model hyperparameter optimization, taking the best performing value from the grid as the optimal parameter set. The spaces searched for GCNN and MPNN models are summarized in Table 3. We also note that this set of hyperparameters is not comprehensive and that the exploration was only over a very modest grid of values. This limited search was necessary to keep computations manageable over the many optimizations performed in this work. However, it is likely that at least somewhat better hyperparameter sets could be found for any particular case with a more complete optimization.

| GCNN Grid Search Space | |
|---|---|
| Batch size | 8*,32 |
| Epochs | 70,100,150*,200,400 |
| Tasks | 1 |
| Convolution layers | [64,64] |
| Dense layers | 128,256,512* |
| Dropout | 0.0, 0.2*, 0.4 |
| Learning rate | 0.005, 0.0005*, 0.001 |

| MPNN Grid Search Space | |
|---|---|
| Batch size | 8*,32 |
| Epochs | 70,100,150,200*,400 |
| Tasks | 1 |
| Atom features | 75 |
| Pair features | 14 |
| T | 1 |
| M | 1 |
| Dropout | 0.0, 0.2*, 0.4 |
| Learning rate | 0.005, 0.0005*, 0.001 |

**Table 3.** Search space for grid search optimization. Definitions of all parameters are given in Refs. [36, 8] (GCNN) and [36, 11] (MPNN). Optimal values used for the model for the full dataset (see section 3.2) are noted with a *.

| | | Type | Chen 14 | Wang 11 | Mathieu 12 | Katrizky 07 | Le 15 | Pan 07 | Pan 13 | Patel 10 | Mean | Stdev | Weighted average | Weighted stdev |
|---|---|---|---|---|---|---|---|---|---|---|---|---|---|---|
| $R^2$ [a] | Original reference | Best result | 0.92 | 0.89 | 0.89 | 0.85 | 0.9 | 0.99 | 0.91 | 0.772 | 0.91 | 0.04 | 0.89 | 0.03 |
| | GCNN | Mean | 0.78 | 0.85 | 0.85 | 0.84 | 0.87 | 0.9 | 0.86 | 0.81 | 0.84 | 0.04 | 0.86 | 0.02 |
| | | Stdev | 0.07 | - | - | 0.03 | - | 0.06 | 0.02 | - | 0.04 | 0.02 | 0.04 | 0.02 |
| | | Diff | -0.14 | -0.04 | -0.04 | -0.01 | -0.03 | -0.09 | -0.05 | 0.04 | -0.05 | 0.05 | -0.03 | 0.02 |
| | | Diff % | -15.1% | -4.6% | -4.6% | -1.6% | -3.7% | -8.9% | -5.8% | 5.2% | -4.9% | 5.8% | -3.7% | 2.37% |
| | MPNN | Mean | 0.83 | 0.87 | 0.87 | 0.87 | 0.86 | 0.96 | 0.86 | 0.82 | 0.87 | 0.04 | 0.86 | 0.01 |
| | | Stdev | 0.06 | - | - | 0.02 | - | 0.02 | 0.07 | - | 0.04 | 0.03 | 0.03 | 0.03 |
| | | Diff | -0.09 | -0.02 | -0.02 | 0.02 | -0.04 | -0.03 | -0.05 | 0.05 | -0.02 | 0.04 | -0.03 | 0.02 |
| | | Diff % | -9.5% | -2.2% | -2.2% | 2.5% | -4.4% | -3.0% | -5.3% | 6.2% | -2.3% | 4.8% | -3.7% | 2.62% |
| MAE (K) [b] | Original reference | Best result | 10.3 | 11.2 | 12 | 13.9 | - | 4.8 | 11.1 | - | 10.5 | 3.1 | 12 | 2.4 |
| | GCNN | Mean | 16.3 | 13.9 | 13.9 | 13 | 18.9 | 9 | 15.1 | 24 | 15.5 | 4.4 | 18.2 | 2.3 |
| | | Stdev | 1.8 | - | - | 1.3 | - | 1.2 | 2.2 | - | 1.6 | 0.4 | 1.5 | 0.4 |
| | | Diff | 6 | 2.8 | 1.9 | -0.9 | - | 4.2 | 4 | - | 3 | 2.3 | 1.7 | 2.8 |
| | | Diff % | 57.9% | 24.9% | 16.1% | -6.3% | -% | 87.5% | 36.1% | -% | 36.0% | 33.0% | 18.8% | 30.1% |
| | MPNN | Mean | 17.8 | 13.3 | 13.3 | 12.2 | 17.9 | 6 | 13 | 24.6 | 14.7 | 5.4 | 17.3 | 2.5 |
| | | Stdev | 2 | - | - | 1.2 | - | 1.2 | 2.2 | - | 1.7 | 0.5 | 1.5 | 0.5 |
| | | Diff | 7.4 | 2.1 | 1.3 | -1.7 | - | 1.2 | 1.9 | - | 2 | 3 | 0.9 | 3.2 |
| | | Diff % | 72.0% | 19.1% | 10.7% | -12.2% | -% | 25.0% | 17.1% | -% | 21.9% | 27.7% | 10.9% | 29.3% |
| RMSE (K) [c] | Original reference | Best result | - | - | - | - | 25.7 | 6.9 | 14.1 | - | 15.6 | 9.5 | 25.3 | 2.9 |
| | GCNN | Mean | 21.5 | 17 | 17 | 19.9 | 27 | 12 | 18.7 | 32.4 | 20.7 | 6.4 | 25.9 | 3.5 |
| | | Stdev | 2.4 | - | - | 2.4 | - | 1.8 | 2.5 | - | 2.3 | 0.3 | 2.4 | 0.2 |
| | | Diff | - | - | - | - | 1.3 | 5.1 | 4.5 | - | 3.7 | 2.1 | 1.4 | 0.7 |
| | | Diff % | -% | -% | -% | -% | 5.1% | 74.6% | 32.2% | -% | 37.3% | 35.1% | 6.4% | 9.1% |
| | MPNN | Mean | 19.8 | 16.2 | 16.2 | 18.1 | 27.7 | 8.4 | 16.9 | 33.6 | 19.6 | 7.7 | 26.2 | 4.3 |
| | | Stdev | 2.9 | - | - | 2.6 | - | 2.1 | 3.8 | - | 2.8 | 0.7 | 2.8 | 0.5 |
| | | Diff | - | - | - | - | 2 | 1.5 | 2.8 | - | 2.1 | 0.6 | 2 | 0.1 |
| | | Diff % | -% | -% | -% | -% | 7.7% | 22.4% | 19.6% | -% | 16.6% | 7.8% | 8.1% | 2.6% |
| AARE(%) [d] | Original reference | Best result | 3.6 | 3.7 | - | - | - | - | 3.2 | - | 3.5 | 0.2 | 3.5 | 0.2 |
| | GCNN | Mean | 5.4 | 4.6 | 4.6 | 4.3 | 4.9 | 3.4 | 4.9 | 6.9 | 4.9 | 1 | 4.9 | 0.4 |
| | | Stdev | 0.8 | - | - | 0.6 | - | 0.6 | 0.8 | - | 0.7 | 0.1 | 0.7 | 0.1 |
| | | Diff | 1.8 | 0.9 | - | - | - | - | 3 | - | 1.9 | 1.1 | 1.9 | 1.0 |
| | | Diff % | 49.1% | 25.7% | -% | -% | -% | -% | 50.3% | -% | 41.7% | 13.9% | 41.5% | 13.1% |
| | MPNN | Mean | 4.8 | 4.4 | 4.4 | 4 | 4.6 | 2.2 | 4.2 | 6.9 | 4.4 | 1.3 | 4.6 | 0.5 |
| | | Stdev | 0.9 | - | - | 0.5 | - | 0.5 | 0.7 | - | 0.6 | 0.2 | 0.6 | 0.2 |
| | | Diff | 1.2 | 0.7 | - | - | - | - | 0.3 | - | 0.7 | 0.4 | 0.7 | 0.4 |
| | | Diff % | 33.3% | 20.2% | -% | -% | -% | -% | 30.2% | -% | 27.9% | 6.9% | 27.8% | 6.5% |

[*] All results are test set results
[a] $R^2$ was also referred to as predictive capability ($Q^2$) [14]
[b] MAE was also referred to as average absolute deviation [22, 23, 25]
[c] RMSE was also referred to as standard error of prediction [21]
[d] AARE was also referred to as average absolute error in percentage [19], average error in percentage [14, 24], and absolute average relative deviation [23].
[e] Wang11 and Mathieu12 used same dataset

**Table 4.** Comparisons between GBDL models and non-boiling point papers (see text for details on the role of boiling point feature). Diff is calculated by (Mean − Original Reference Best Result), and Diff % is calculated by $\left(\frac{\text{Mean} - \text{Original Reference Best Result}}{\text{Original Reference Best Result}}\right) * 100$.

| | | Type | Carroll 10 | Carroll 11 | Carroll 15 | Godinho 11 | Mean | Stdev | Weighted mean | Weighted stdev |
|---|---|---|---|---|---|---|---|---|---|---|
| $R^2$ | Original reference | Best result | 0.99 | 0.99 | 1 | 0.97 | 0.99 | 0.01 | 0.99 | 0.01 |
| | GCNN | Mean | 0.97 | 0.93 | 0.98 | 0.74 | 0.9 | 0.11 | 0.93 | 0.06 |
| | | Stdev | 0.01 | 0.03 | - | 0.21 | 0.08 | 0.11 | 0.04 | 0.05 |
| | | Diff | -0.02 | -0.07 | -0.02 | -0.24 | -0.08 | 0.1 | -0.07 | 0.05 |
| | | Diff % | -1.50% | -6.70% | -1.90% | -24.40% | -8.60% | 10.80% | -6.3% | 5.5% |
| | MPNN | Mean | 0.98 | 0.96 | 0.98 | 0.96 | 0.97 | 0.01 | 0.97 | 0.01 |
| | | Stdev | 0 | 0.02 | - | 0.05 | 0.02 | 0.02 | 0.02 | 0.01 |
| | | Diff | -0.01 | -0.03 | -0.02 | -0.01 | -0.02 | 0.01 | -0.02 | 0.01 |
| | | Diff % | -1.00% | -3.30% | -1.90% | -1.40% | -1.90% | 1.00% | -2.6% | 1.1% |
| MAE (K) [a] | Original reference | Best result | 2.9 | 2.5 | 2.2 | 2.8 | 2.6 | 0.31 | 2.6 | 0.2 |
| | GCNN | Mean | 6.4 | 7.3 | 5.1 | 6 | 6.2 | 0.93 | 6.9 | 0.7 |
| | | Stdev | 2.8 | 0.4 | - | 1 | 1.41 | 1.26 | 1.0 | 1.2 |
| | | Diff | 3.5 | 4.8 | 2.9 | 3.2 | 3.6 | 0.85 | 4.3 | 0.8 |
| | | Diff % | 121.90% | 192.00% | 130.50% | 114.30% | 139.70% | 35.50% | 169.4% | 36.6% |
| | MPNN | Mean | 4.8 | 5.9 | 3.9 | 2.5 | 4.27 | 1.45 | 5.4 | 1.0 |
| | | Stdev | 0.6 | 1 | - | 0.5 | 0.7 | 0.27 | 0.9 | 0.2 |
| | | Diff | 1.9 | 3.4 | 1.7 | -0.3 | 1.67 | 1.53 | 2.8 | 1.1 |
| | | Diff % | 64.50% | 136.80% | 77.30% | -11.50% | 66.80% | 60.90% | 110.1% | 47.2% |
| RMSE (K) | Original reference | Best result | - | - | - | - | - | - | - | - |
| | GCNN | Mean | 8.3 | 11.4 | 6.1 | 8.3 | 8.5 | 2.19 | 10.3 | 1.9 |
| | | Stdev | 2.6 | 1.9 | - | 1.2 | 1.87 | 0.7 | 2.0 | 0.4 |
| | | Diff | - | -% | - | - | - | - | - | - |
| | | Diff % | -% | -% | -% | -% | -% | -% | -% | - |
| | MPNN | Mean | 6.3 | 8.7 | 6.5 | 3.2 | 6.15 | 2.27 | 7.8 | 1.7 |
| | | Stdev | 0.9 | 1.8 | - | 0.5 | 1.08 | 0.66 | 1.5 | 0.5 |
| | | Diff | - | -% | - | - | - | -% | - | - |
| | | Diff % | -% | -% | -% | -% | -% | -% | -% | - |
| AARE (%) | Original reference | Best result | - | 0.8 | - | - | 0.8 | -% | 0.8 | - |
| | GCNN | Mean | 2.2 | 2.3 | 1.6 | 1.8 | 1.98 | 0.34 | 2.2 | 0.2 |
| | | Stdev | 1 | 1.3 | - | 0.3 | 0.88 | 0.52 | 1.2 | 0.3 |
| | | Diff | - | 1.5 | -% | -% | 1.53 | - | 1.5 | - |
| | | Diff % | -% | 191.10% | -% | -% | 191.10% | -% | 191.1% | - |
| | MPNN | Mean | 1.6 | 1.9 | 1.2 | 0.8 | 1.37 | 0.5 | 1.7 | 0.3 |
| | | Stdev | 0.2 | 0.3 | - | 0.1 | 0.22 | 0.09 | 0.3 | 0.1 |
| | | Diff | - | 1.1 | -% | -% | 1.1 | - | 1.1 | - |
| | | Diff % | -% | 137.50% | -% | -% | 137.50% | -% | 137.5% | - |

[*] All results are test set results
[a] MAE was also referred to as average absolute deviation [15, 16, 18, 20].

**Table 5.** Comparisons between GBDL models and boiling point papers (see text for details on role of boiling point feature). Diff is calculated by (Mean − Original Reference Best Result) and Diff % is calculated by $\frac{\text{Mean} - \text{Original Reference Best Result}}{\text{Original Reference Best Result}}$

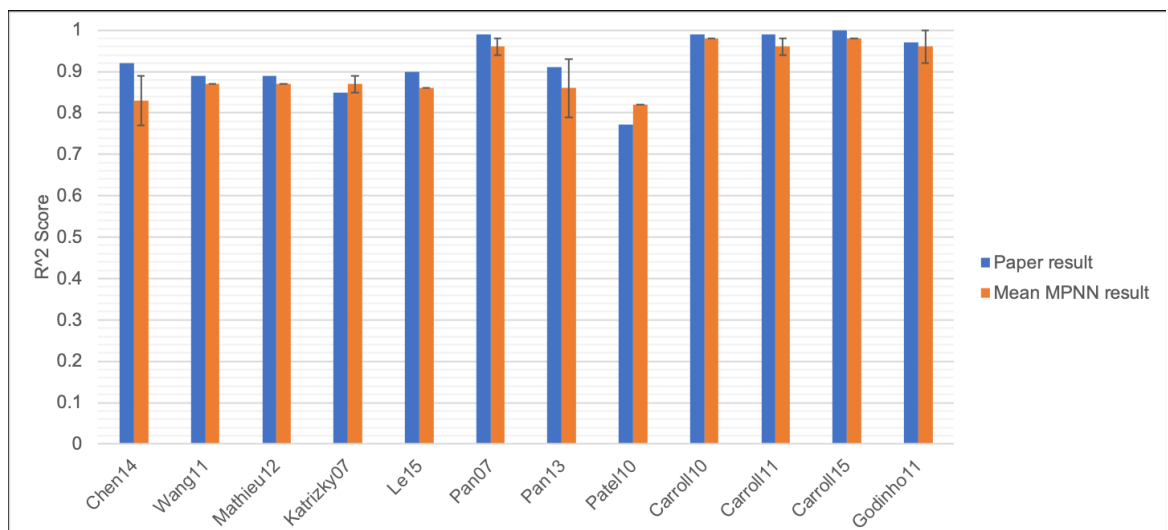

(A)

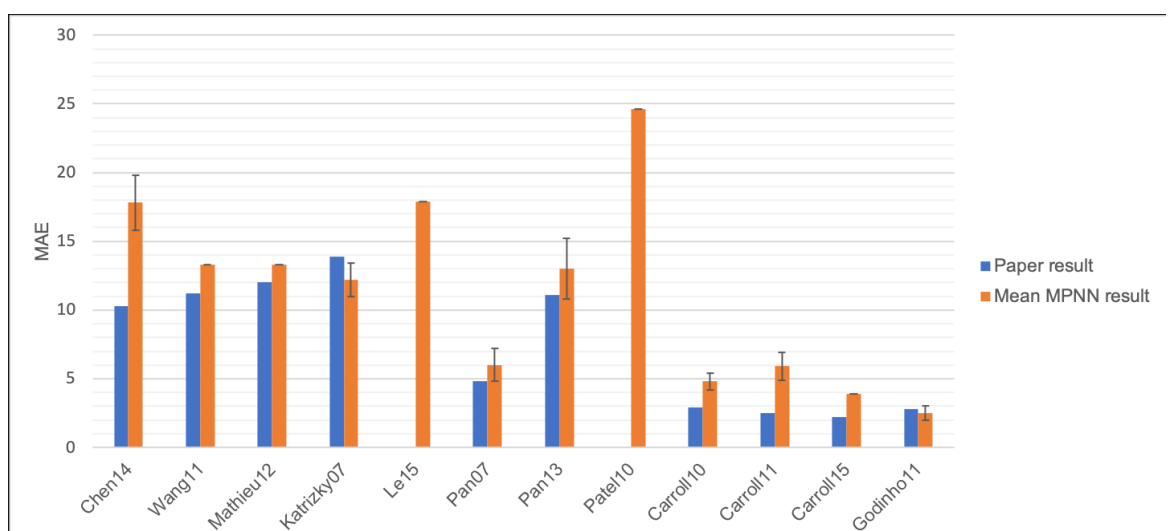

(B)

*Le15 and Patel10 didn't provide MAE.
*Wang11, Mathieu12, Le15, Patel 10, and Carroll15 were only tested on given test set once. More details please see Sec 2.4

Figure 2. (A) MPNN $R^2$ result comparisons. (B) MPNN MAE result comparisons. Error bars represent one standard deviation of values over all validation or test sets studied (see section 3.1 text for discussion of validation and test sets).

## 3 Results and Discussion

### 3.1 Comparison to Literature Results

The developed GCNN and MPNN models were compared against the results of a number of previous investigations into flash point prediction. We excluded testing against previous papers that made use of the DIPPR dataset. This choice was motivated by two factors. First, DIPPR contains both experimental and model data, the latter predicted by data centric approaches. Some authors did not clearly state that they were including only experimental data, and if they included model data it might skew their models and would not be a proper dataset for testing machine learning approaches. Second, because DIPPR is proprietary, authors did not share their datasets (or at least should not have shared sets if they did), which means that in almost all cases we cannot determine exactly which data points were used in the fitting. Excluding comparisons to DIPPR studies still allowed comparisons to a number of other studies without such issues. We divide the analysis into two categories: comparisons with papers that did not use boiling point as a molecular feature (non-boiling point papers) and comparisons with papers that did use boiling point as a feature (boiling point papers). We make this separation because, while using measured physical properties like boiling point can boost prediction accuracy, getting such features for any new system requires synthesizing that system and then measuring the property. These steps are enormously more difficult for new systems than using features that can be automatically generated from just the molecular description. Therefore, boiling point papers represent a qualitatively different type of model than non-boiling point papers and should be compared separately. Details regarding the results of comparisons are recorded in Table 4 and Table 5. In Table 4 and Table 5, we provide both a uniform weighted (a simple mean) and dataset relative size weighted average score for each metric. The uniform weighted approach simply weights the value from each paper by 1/(number of papers) and represents a typical value you might get for a given paper dataset. The dataset relative size weighted approach weights the value from each paper by (dataset size for that paper)/(sum of all dataset sizes) and represents a typical value you might get for a random sample from all the available data. Dataset sizes are taken from the "Size" row in Table 1. For each weighted average we also provide the corresponding weighted standard deviation. Equations used to calculate weighted average and weighted standard deviation are shown in Equation (5) and (6) respectively.

$$\sum_{i=1}^{N}(w_i * x_i) \quad (5)$$

$$\sqrt{\frac{\sum_{i=1}^{N} w_i (x_i - \overline{x_w})^2}{\frac{(N'-1) \sum_{i=1}^{N} w_i}{N'}}} \quad (6)$$

\* N is the number of datasets involved in the calculation; N' is the number of non-zero weights; $w_i$ are weights; $\overline{x_w}$ is weighted average.

For this paper, N' and N are the same and the sum of weights is 1 in Equation (6).

As the goal for this paper is to assess GBDL models by comparing with specific previous QSPR studies, our discussion will be based on mean values. Mean values from Table 4 are also shown in bar charts comparing MPNN results with previous studies for two of the most widely used statistics, $R^2$ and MAE, in Figure 2A and 2B, respectively. Recall that we are using statistically similar test data and comparison in most cases, and exact test data comparisons for some cases (see Methods section). Upon analysis of the comparisons, it becomes immediately clear that the MPNN model outperforms the GCNN model as MPNN has better average scores, and generally better scores for each dataset, for all four metrics we are using to evaluate models. Given these observations, further discussion of comparisons with paper results will only consider the MPNN model.

With respect to the non-boiling point papers, the MPNN model generally performed worse than the average reported results of the papers. We obtain decreases in $R^2$ of 2.3% on average, and an average increase in MAE of 21.9% (2.0 K). Across each metric, significant improvements were observed in comparisons to two papers. Specifically, the MAE and $R^2$ values for katritzky07[28] dropped by 12.2% and increased by 2.5% respectively, and the $R^2$ value for patel10 increased by 6.2%. Performance worsened significantly in only one instance. The MAE and $R^2$ values for chen14[19] increased by 72% and decreased by 9.5%. In all other cases, the MPNN model performance was essentially comparable to the results given in the paper.

When comparing our MPNN model with boiling point papers (Table 5) we obtain only modest decreases in $R^2$ of 1.9% on average, and an average increase in MAE of 66.8% (1.7 K). This increase is a large percentage, although still only a few degrees K. We believe that the larger errors from the MPNN vs. models in these papers is due to their use of the boiling point feature which, as mentioned above, enhances accuracy, but at the cost of greatly increasing the amount of work needed to predict a new compound.

Here we consider in more detail the comparison to the non-boiling point studies. While the MPNN model does underperform on average when compared to previous studies, the results are sometimes better, often similar, and always quite close even when worse. Evidence of this can be seen across all comparisons in the absolute errors. On average, these errors are within 1—4 K (and differences in the MAE and RMSE are always less than 3.5 K and 2 K, respectively) which is overall a relatively small discrepancy compared to the mean or range of flash point values (hundreds of degrees), standard deviations in the datasets (37-74 K), and even differences between different experiments on the same material, which can often be one or two K. This suggests that MPNN, with just the modest level of optimization pursued here, can already obtain performance only slightly worse on average than comparable models built from traditional features. This result suggests that traditional feature approaches may be better for many problems, but that GBDL can yield similar performance and is worth considering if easier to implement. Furthermore, the GBDL results here are close enough to those from the previous work that it is possible that more extensive optimization of the present approaches, and/or additional methodological improvements, will make these GBDL approaches superior to the more traditional ones. These results are somewhat in contrast to those presented by some previous references, which showed GCNN[6,7,9,10] and MPNN[11] outperforming previous QSPR studies on datasets related to toxicity, solubility, and quantum chemistry predicted properties. The origin of the greater success of GBDL vs. traditional QSPR approaches in previous assessments compared to this one is not clear but may be related to more extensive optimization of the GBDL networks for those problems, or some aspects of the datasets being studied. In particular, the datasets used here are all either the same size or smaller than those used in the original papers (see Table 1 for sizes) because we could not convert some compounds to SMILES strings (see section 2.2 for discussion).

Having less data might lead to worse models and could create reductions in our average performance. However, we don't see any correlation of our model performance vs. the original papers and the amount of data lost. For example, the largest percent reduction between the original dataset and our dataset is the 27% reduction for Godinho11, but the average MAE score of our MPNN model outperforms the original paper's MAE score by 0.3 K. More work is clearly needed to understand when these GBDL methods are likely to be most effective compared to traditional approaches.

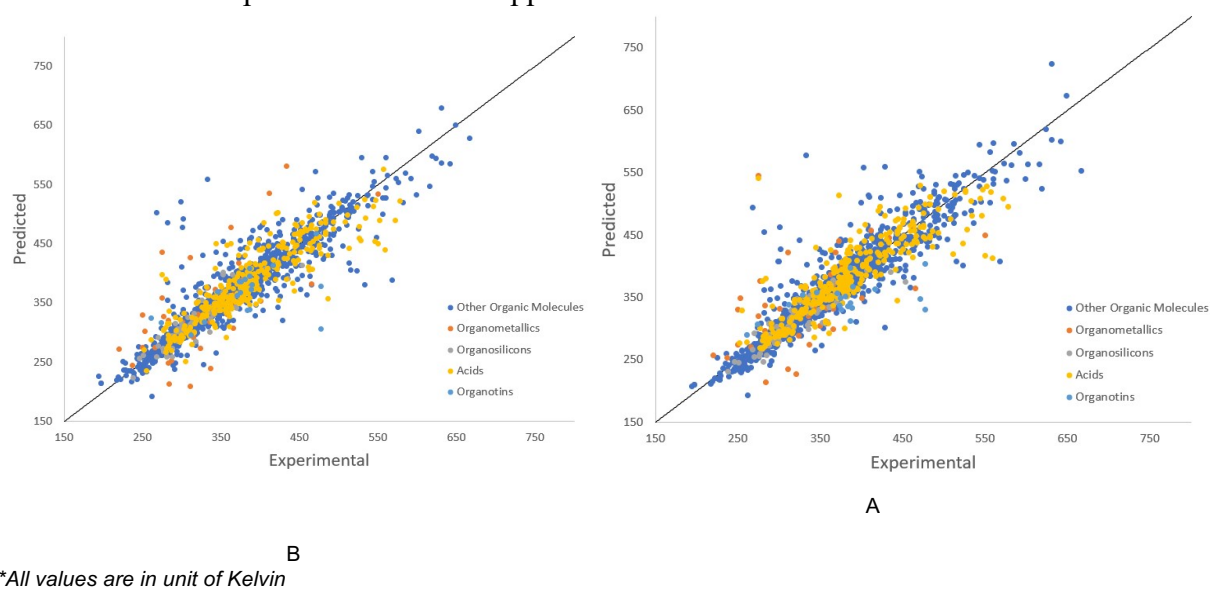

*All values are in unit of Kelvin

**Figure 3.** (A) Predicted vs. experimental values of flash point on full integrated distribution test set for MPNN. (B) Predicted vs. experimental values of flash point on full integrated distribution test set for GCNN.

|  |  | Full | Acids | Organometallics | Organogermaniums | Organosilicons | Organotins |
|---|---|---|---|---|---|---|---|
| Test set info | Mean (K) | 365.36 | 362.38 | 323.08 | 326.76 | 316.24 | 405.35 |
|  | Std | 71.64 | 60.94 | 58.80 | 40.28 | 46.50 | 75.94 |
|  | Size | 1621 | 239 | 290 | 18 | 240 | 25 |
| RMSE (K) | GCNN | 30.79 | 27.82 | 37.81 | 22.82 | 19.90 | 64.17 |
|  | MPNN | 30.55 | 29.87 | 37.82 | 17.34 | 20.92 | 74.87 |
| MAE (K) | GCNN | 19.14 | 19.48 | 22.03 | 18.90 | 13.15 | 45.47 |
|  | MPNN | 18.76 | 20.56 | 20.14 | 14.61 | 12.65 | 42.57 |
| $R^2$ | GCNN | 0.82 | 0.79 | 0.61 | 0.91 | 0.82 | 0.51 |
|  | MPNN | 0.83 | 0.77 | 0.64 | 0.82 | 0.81 | 0.18 |
| AARE (%) | GCNN | 5.21 | 5.28 | 6.76 | 5.57 | 4.22 | 10.29 |
|  | MPNN | 5.07 | 5.65 | 6.21 | 4.51 | 4.04 | 9.32 |

* All results are test set results

**Table 6.** Statistics and results for model fits on molecule type subsets.

**3.2 Performance of Models on Entire Dataset**

As noted in section 2.1, in developing our comprehensive data for the testing set, we included DIPPR, and therefore also included some studies that had potentially useful datasets that could expand our collection, even if those studies used DIPPR for some or most of their data.

The prediction performance of GCNN and MPNN models on the full dataset are presented in parity plots in Figure 3 and detailed statistics regarding the test dataset and test results are in Table 6. Unlike in the fits to specific papers, which showed MPNN clearly superior to GCNN, here both models gave essentially identical results. In general, the GBDL on the complete dataset is less accurate than on any of the subsets used in the specific papers discussed above. This is likely due to a larger and more complex database being used here compared to most other studies.

To test the robustness of GCNN and MPNN on our entire dataset, we did a second full fit for both models, starting again from a different random initialization of the weights in the neural networks. The difference of MAE and $R^2$ of these two tests were 0.73 K and 0.005, respectively, for GCNN and 0.49 K and 0.007, respectively, for MPNN. These changes in MAE and $R^2$ are less than 4% and 0.85%, respectively, and therefore suggest that our results will not change significantly by simply refitting from a different starting point.

**3.3 Statistics and Results for Model Fits on Molecule Type Subsets**

It is of some interest to ask if the overall model might perform differently for different chemical classes, an effect which can be readily explored in our large heterogeneous database. The chemical classes were discussed in the Methods section and details regarding chemical test datasets and associated prediction statistics are shown in Table 6.

Both the GCNN and MPNN models were optimized on the integrated dataset and the fit accuracy assessed on subset datasets from each chemical compound group. As observed in the integrated dataset test, both models performed about equivalently. MPNN outperformed GCNN by a significant margin in only the organogermanium dataset, where it is better on all statistics, but underperforms by a significant margin for RMSE and $R^2$ on the organotin dataset (but is actually outperforming GCNN for MAE and AARE). In this discussion we therefore report both values, with MPNN values before GCNN values, with the latter immediately following in parentheses. We focus on RMSE as a representative statistic except when other statistics suggest a clearly different conclusion. The acids dataset distribution matched that of the full dataset quite well. Predictions resulted in an RMSE of 29.87 K (27.82 K), slightly better than the performance of the full dataset. The organometallic dataset exhibited a poorer RMSE of 37.82 K (37.81 K). The low $R^2$ value of 0.64 (0.61) is likely due to a number of outliers throughout the organometallic dataset, consistent with the difference in RMSE from the full dataset. Our GBDL models predicted both the organogermaniums and organosilicon compounds well, achieving RMSE values of 17.34 K (22.82 K) and 20.92 K (19.90 K) respectively. The corresponding $R^2$ values of 0.82 (0.91) (organogermaniums) and 0.81 (0.82) (organosilicons) show that the model was capable of fitting to the overall data distribution while robustly modeling these chemical subclasses. The organotin compounds proved to be a challenge, with a RMSE of 74.87 K (64.17 K) and $R^2$ of 0.18 (0.51). The fact that the RMSE is comparable to the standard deviation of the data is a sign that the model is capturing little of flash point physics for this dataset. In particular, we note that a model that simply gives the mean of a dataset will achieve an RMSE equal to the dataset's standard deviation, so these errors suggest that the model is doing little better than just guessing the mean of the data. The reduced accuracy of the model for some of these materials suggests that they may be meaningfully distinct from the overall dataset, and that there were only a small number in the overall dataset to guide the fitting for this chemistry. While it would be valuable to understand to what extent the GBDL models' errors are tied to specific aspects of the chemistry in each dataset, we were not able to extract any such understanding from the trends we observed. While such correlations may exist, they are very difficult to separate from the many other numerical factors controlling the results of these models. However, these tests do show that the optimized MPNN and GCNN models can perform somewhat unreliably when considering specific chemical domains of interest, and that careful testing and possible refinement is advisable before applying these models to specific subsets of data.

**4 Conclusions**

This work conducted comparisons between two GBDL models (MPNN and GCNN) and previous QSPR studies in predicting flash point. As the comparison results show in Table 4 and Table 5, in general the GBDL models do not outperform traditional QSPR methods for the datasets studied here. More specifically, we found that MPNN is more accurate than GCNN and that MPNN gives results comparable to, although slightly worse on average than, traditional QSPR methods. This result suggests that either the essential physics for flash points is already present in the traditional QSPR features or that we have inadequate data and/or optimization to training the GBDL approaches. Regarding the latter issue, the limited hyperparameter optimization of the MPNN models suggests that further efforts could yield results as good as or better than traditional methods, although additional study is needed to support such a claim. We assembled the largest flash point dataset to date, containing 10575 unique molecules, and found that the MPNN model gave an overall RMSE, MAE, $R^2$, and AARE of 30.55 K, 18.76 K, 0.83, and 5.07%, respectively, providing a generally accurate model trained on a wide range of data. However, tests on five chemical subsets extracted from this integrated dataset showed predictions were significantly worse for some of these chemistries. This could be related to the small amount of data in these subsets and/or some aspects of their chemistry. Given this result, care should be taken in use of the model in specific chemical subdomains. In future work it would be of interest to see if more extensive hyperparameter optimization could improve GBDL's flash point performance, and if a model with better results in targeted chemistry domains with limited data (e. g., different organometallics) could be established.

## 5 Data and Code Dissemination

We have shared the following data and code information:

1. The software used to manage the workflow of calling and testing the models. The code can be found at https://github.com/uw-cmg/MoleProp. This repository also contains training and test sets for each fold for the optimizations done for the paper comparisons and all information on the final optimized hyperparameters for each paper comparison.

2. For each model testing run we have saved the computed metrics after each cross-validation fold represented in the test set. For the total dataset studies, we have also saved the specific test sets (i. e. the integrated dataset test set and the chemical subset test sets). These are shared in the following files and folders along with a README file for guidance. Note that the proprietary DIPPR data has been removed from this data. For "explicit test" datasets (see section 2.4 for details on explicit test dataset selection), we have marked the training data and test data in the dataset we provide in the supplementary materials. All other data points used in paper comparisons are used as both training and test data through cross validation and are marked as such in the the dataset we provide in the supplementary materials.

- Plots file that contains parity plots and residual histogram plots for each test fold for the cross-validation tests associated with paper comparisons
- Outliers file for every test dataset considered in the study. These outliers are defined as predicted values with errors compared to true values of greater than 100 K (in files outliers.csv). Note that these outliers are provided for convenience and we do not mean to imply that they are in some way incorrect data, although that might be the case. Note that, consistent with our removal of all proprietary DIPPR database data, outliers that were from the DIPPR database have been removed.
- Final test results for each paper comparison test
- Digital excel files with data used for Fig 2A, 2B.

All of the items in 2 and 3 have been submitted as part of Supplement Information as well as shared in the form of tarballs on Figshare at 10.6084/m9.figshare.9275210.

**Acknowledgements**

This research was primarily supported by NSF through the University of Wisconsin Materials Research Science and Engineering Center (DMR-1720415). Computations for this work were performed with support from the University of Wisconsin Center for High-Throughput Computing and Wisconsin Applied Computing Center. The authors declare no conflicts of interest.

[1]. Vidal, M., Rogers, W. J., Holste, J. C., & Mannan, M.S. (2004). A Review of Estimation Methods for Flash Points and Flammability Limits, *23*(1), 47--55.

[2]. Nieto-draghi, C., Fayet, G., Creton, B., Rozanska, X., Rotureau, P., Hemptinne, J. De, Adamo, C. (2015). A General Guidebook for the Theoretical Prediction of Physicochemical Properties of Chemicals for Regulatory Purposes.

[3]. Phoon, L. Y., Azri, A., Hashim, H., & Mat, R. (2014). A Review of Flash Point Prediction Models for Flammable Liquid Mixtures.

[4]. Liu, X., & Liu, Z. (2010). Research Progress on Fash Point Prediction. *Journal of Chemical & Engineering Data*, *55*(9), 2943--2950.

[5]. Oriol Vinyals, Samy Bengio, M. K. (2016). Ordermatters: Sequence To Sequence for Sets. *Iclr*.

[6]. Kearnes, S., McCloskey, K., Berndl, M., Pande, V., & Riley, P. (2016). Molecular graph convolutions: moving beyond fingerprints. *Journal of Computer-Aided Molecular Design*, *30*(8), 595--608.

[7]. Korolev, V., Mitrofanov, A., Korotcov, A., & Tkachenko, V. (2019). Graph convolutional neural networks as "general-purpose" property predictors: the universality and limits of applicability, 1--8.

[8]. Altae-Tran, H., Ramsundar, B., Pappu, A. S., & Pande, V. (2017). Low Data Drug Discovery with One-Shot Learning. *ACS Central Science*, *3*(4), 283--293.

[9]. Coley, C. W., Jin, W., Rogers, L., Jamison, T. F., Jaakkola, T. S., Green, W. H., … Jensen, K. F. (2019). A graph-convolutional neural network model for the prediction of chemical reactivity. *Chemical Science*, *10*(2), 370--377.

[10]. Zeng, M., Kumar, J. N., Zeng, Z., Savitha, R., Chandrasekhar, V. R., & Hippalgaonkar, K. (2018). Graph Convolutional Neural Networks for Polymers Property Prediction

[11]. Gilmer, J., Schoenholz, S. S., Riley, P. F., Vinyals, O., & Dahl, G. E. (2017). Neural Message Passing for Quantum Chemistry.

[12]. Project 801, evaluated process design data, public release documentation, design institute for physical properties (DIPPR), AIChE, 2015.

[13]. Ramsundar, B., Eastman, P., Walters, P., Pande, V., Leswing, K., & Wu, Z. (2019). *Deep Learning for the Life Sciences*. O'Reilly Media.

[14]. Wang, G.-B., Chen, C.-C., Liaw, H.-J., & Tsai, Y.-J. (2011). Prediction of Flash Points of Organosilicon Compounds by Structure Group Contribution Approach. *Industrial & Engineering Chemistry Research*, *50*(22), 12790--12796.

[15]. Carroll, F. A., Lin, C. Y., & Quina, F. H. (2011). Simple method to evaluate and to predict flash points of organic compounds. *Industrial and Engineering Chemistry Research*, *50*(8), 4796--4800.

[16]. Carroll, F. A., Lin, C.-Y., & Quina, F. H. (2010). Improved Prediction of Hydrocarbon Flash Points from Boiling Point Data. *Energy & Fuels*, *24*(9), 4854--4856.

[17]. Carroll, F. A., Lin, C.-Y., & Quina, F. H. (2010). Calculating Flash Point Numbers from Molecular Structure: An Improved Method for Predicting the Flash Points of Acyclic Alkanes. *Energy & Fuels*, *24*(1), 392--395.

[18]. Carroll, F. A., Brown, D. M., & Quina, F. H. (2015). Predicting Boiling Points and Flash Points of Monochloroalkanes from Structure. *Industrial & Engineering Chemistry Research*, *54*(1), 560--564.

[19]. Chen, C. P., Chen, C. C., & Chen, H. F. (2014). Predicting flash point of organosilicon compounds using quantitative structure activity relationship approach. Journal of

[20]. Godinho, J. M., Lin, C.-Y., Carroll, F. A., & Quina, F. H. (2011). Group Contribution Method To Predict Boiling Points and Flash Points of Alkylbenzenes. *Energy & Fuels*, *25*(11), 4972--4976.

[21]. Le, T. C., Ballard, M., Casey, P., Liu, M. S., & Winkler, D. A. (2015). Illuminating Flash Point: Comprehensive Prediction Models. Molecular Informatics, 34(1), 18--27.

[22]. Pan, Y., Jiang, J., Zhao, J., & Wang, R. (2007). QSPR studies for predicting flash points of alcohols using group bond contribution method with back-propagation neural networks. In P. Jing, G and Gao, J and Zhou, A and Gou (Ed.), *Progress in Mining Science and Safety Technology, Pts A and B* (pp. 1237--1244).

[23]. Mathieu, D. (2012). Flash Points of Organosilicon Compounds: How Data for Alkanes Combined with Custom Additive Fragments Can Expedite the Development of Predictive Models. *Industrial & Engineering Chemistry Research*, *51*(43), 14309--14315.

[24]. Pan, Y., Jiang, J., Wang, R., Zhu, X., & Zhang, Y. (2013). A novel method for predicting the flash points of organosilicon compounds from molecular structures. Fire and Materials

[25]. Patel, S. J., Ng, D., & Mannan, M. S. (2010). QSPR Flash Point Prediction of Solvents Using Topological Indices for Application in Computer Aided Molecular Design (vol 48, pg 7378, 2009). *Industrial & Engineering Chemistry Research*, 49(17), 8282--8287.

[26]. Saldana, D. A., Starck, L., Mougin, P., Rousseau, B., Pidol, L., Jeuland, N., & Creton, B. (2011). Flash point and cetane number predictions for fuel compounds using quantitative structure property relationship (QSPR) methods. *Energy and Fuels*, 25(9), 3900--3908 .

[27]. Mathieu, D., & Alaime, T. (2014). Insight into the contribution of individual functional groups to the flash point of organic compounds. *Journal of Hazardous Materials*, *267*, 169--174.

[28]. Katritzky, A. R., Stoyanova-Slavova, I. B., Dobchev, D. A., & Karelson, M. (2007). QSPR modeling of flash points: An update. *Journal of Molecular Graphics and Modelling*, *26*(2), 529--536.

[29]. Kim, S., Chen, J., Cheng, T., Gindulyte, A., He, J., He, S., … Bolton, E. E. (2019). PubChem 2019 update: improved access to chemical data. Nucleic acids research, 47(D1), D1102–D1109.

[30]. Silane, Silicone & Metal-Organic Materials Innovation: Gelest. (n.d.).

[31]. Dean, J. A., & Lange, N. A. (1999). *Lange's handbook of chemistry*. New York: McGraw-Hill.

[32]. Carson, P. A. (2002). *Hazardous Chemicals Handbook.* Oxford: Butterworth-Heinemann.

[33]. Aristarán, M., Tigas, M., & Merrill, J. B. (2018, June 04). Tabula (Version 1.2.1) [Computer software].

[34]. Lowe, D. M., Corbett, P. T., Murray-Rust, P., & Glen, R. C. (2011). Chemical Name to Structure: OPSIN, an Open Source Solution. *Journal of Chemical Information and Modeling*, *51*(3), 739--753.

[35]. RDKit: Open-source cheminformatics; http://www.rdkit.org

[36]. Wu, Z., Ramsundar, B., Feinberg, E. N., Gomes, J., Geniesse, C., Pappu, A. S., … Pande, V. (2018). MoleculeNet: A benchmark for molecular machine learning. *Chemical Science*.